\documentclass[aip,twocolumn,letterpaper]{revtex4}

\usepackage{epsfig}
\usepackage{xcolor}

\usepackage{graphics}
\usepackage{epsfig}

\usepackage[margin=1.0in]{geometry}

\begin{document}
\title{Non-Ambipolarity of Microturbulent Transport} 
\author{Allen H Boozer}
\affiliation{Columbia University, New York, NY  10027 \linebreak ahb17@columbia.edu}

\begin{abstract} 

When exact magnetic surfaces are assumed to exist, the gyrokinetic theory of microturbulence gives the same radial transport for ions and electrons.  But, exact magnetic surfaces do not exist in the presence of what is called electrostatic microturbulence.  When the plasma pressure is non-zero, a turbulent electric potential is accompanied by a  turbulent magnetic field, which splits the rational magnetic surfaces with which it resonates.  If the magnetic field is assumed to have an ideal topology-conserving evolution, delta function current densities arise on resonant surfaces.  The singularity of the current density allows islands to open quickly, but there is no singularity that allows a rapid closure.  Islands remain and do not flutter into and out of existence.  A relative rotation of the electron fluid in neighboring island chains produces a non-dissipative force that can lock the islands together  and produce non-ambipolar transport.  At sufficient plasma pressure, the islands associated with different resonant rational surfaces can overlap.  When this occurs some magnetic field lines will cross the entire radial region occupied by overlapping islands.  The effect on the electron fluid is to create a viscosity-like force, which is dissipative and tends to remove gradients in the electron rotation.  This also produces non-ambipolar transport.  Under many assumptions, the island locking force is larger than the viscosity-like force. 

\end{abstract}

\date{\today} 
\maketitle


\section{Introduction}

Nevins, Wang, and Candy \cite{Nevins:2011},  Connor, Hastie, and Zocco \cite{Connor:2013}, and Terry et al \cite{Terry:2015} have discussed magnetic surface breakup produced by what are called electrostatic instabilities, primarily the ion temperature gradient (ITG) instability.  The focus has been on the enhancement of the electron heat transport, which is not large when the plasma $\beta\equiv 2\mu_0p/B^2$ is small.  


An effect that can arise before the electron heat transport is appreciable is a modification of the radial electric field  that is required to preserve quasi-neutrality in stellarators. \color{black}   In optimized stellarators, the difference in the electron $\Gamma_e$ and the ion particle flux $\Gamma_i$ is only a small fraction, $f_{na}$, of the thermal flux, which is predominately microturbulent;
\begin{equation}
f_{na}\equiv \frac{|\Gamma_e -  \Gamma_i|a}{n D_T},
\end{equation}
where $D_T$ is the thermal diffusion coefficient, $a$ is the plasma radius, and $n$ is the number density of electrons. \color{black} 

A modification in the radial electric field is of great practical importance because neoclassical transport in stellarators usually gives more rapid ion than electron transport, which requires a radial electric field that confines ions---particularly high charge state impurities.  Enhancing just the electron transport by the factor $f_{na}$ would reverse the sign of the electric field and expel impurities.   A large expulsion of impurities in W7-X experiments was found \cite{W7-X:2023} when the microturbulent transport exceeded the expected neoclassical, which requires an explanation.  

Helander and Simakov \cite{Helander:2008} discussed the lack of effect of electrostatic microturbulence on the large scale radial electric field.  The turbulent Reynolds stress can produce short-scale zonal flows that tend to reduce the microturbulent transport. They did not discuss the effect on electron transport of the breaking of the magnetic surfaces by the turbulence.  Their arguments imply the non-ambipolar radial transport of the ions is entirely due to neoclassical, not turbulent, transport effects.   

Even within neoclassical theory, the intrinsic difference in the electron and ion transport can be made very small, and the microturbulent transport \color{black}  in stellarator and tokamak experiments tends to be much larger than the neoclassical.   Consequently, $f_{na}$ will be defined relative to the typical gyro-Bohm radial energy transport seen in experiments;
\begin{eqnarray}
D_T \approx D_{gB}; \\
D_{gB} \equiv \frac{\rho_s}{a} \frac{T}{eB},
\end{eqnarray}  
where $\rho_s$ is the ion gyroradius.  \color{black} The implication is that $f_{na}$ is very small compared to unity; $10^{-3}$ is not an unreasonable value.

This paper points out two physical effects due to electrostatic instabilities that enhance the electron transport and gives qualitative estimates of their effects.   The qualitative estimates should allow microturblent simulations to obtain quantitative information.

Section \ref{sec:B-pert} shows that when the phase velocity of electric potential fluctuations are of order the speed of sound $C_s=\sqrt{T/m_i}$ then magnetic perturbations are produced that are comparable to the plasma $\beta$ times the electric potential perturbations divided by $T/e$, Equation (\ref{Phi/T}).  

The non-ideal terms that allow the breaking of magnetic field lines are intrinsically very small.  As explained in Section \ref{sec:islands}, magnetic field lines can still form magnetic islands and become chaotic for two reasons. 

\begin{enumerate}

\item  The response of an ideally evolving magnetic field to a perturbation that has a component that is resonant and normal to a rational surface produces a delta function current on that surface \cite{Current-sing}.  No matter how small the non-ideal terms such as resistivity may be, an island quickly opens, but there is no current density singularity that allows a fast closure.  The islands at resonant rational surfaces remain and do not flutter into and out of existence.

The far greater inertia of the ion fluid relative to the electron allows ions to depart an approximate distance $\rho_s\equiv C_s/(eB/m_i)$ from the magnetic field lines.  An electric potential arises to bring the electrons along and preserve quasi-neutrality.  Indeed, it is the disparate response of ions and electrons that allows electrostatic microturbulence.  An implication is that the electron fluid is tied to magnetic structures, such as magnetic islands, in a way that ions are not, especially since the width of the islands of primary relevance is comparable to $\rho_s$.

As shown in Section \ref{sec:separated}, the island-associated current, which can be approximated as a delta function, on one rational surface, crossed with the magnetic field associated with an island on a neighboring rational surface gives a force with a sinusoidal dependence on the relative phases of the two islands.  This force is not dissipative just as a road with sinusoidal hill and valleys is not dissipative.  \color{black} Nonetheless, a car even in the neutral gear is dissipative and will come to rest in one of the valleys, even when a force that is too small to push it over a hill is applied to the car.  Similarly, two islands will lock together when the sinusoidal force between them becomes sufficiently strong compared to the force, Section \ref{sec:NA}, associated with the intrinsic non-ambipolarity of $f_{na}$ or with neutral beam injection.   Such forces tend to maintain a shear in the plasma rotation and determine the phase at which the islands lock.  The effect is well known in the theory \color{black} of the interaction of non-axisymmetric magnetic fields with plasmas \cite{Boozer:NF},  of tearing modes \cite{Fitzpatrick}, and \color{black} tokamak disruptions \cite{Buttery:1999} in which the islands have low poloidal mode numbers, such as $m=1$ to 3, but the same physics holds for the high mode numbers of the magnetic islands that arise in microturbulence, $m\approx a/\rho_s$, where $a$ is the plasma radius.  

A locking together of the rotation of the various magnetic surfaces impedes the electron flow from carrying the diamagnetic current.  This changes the radial electric field and the confinement of impurities.


\item  A second microturbulent effect causes a viscosity-like dissipation of the flow of the electrons.  This enhances the electron radial transport, which modifies the radial electric field.  This modification implies momentum transport.  

Unlike symmetry breaking by an external magnetic field, in which momentum changes are balanced by forces on the coils, non-ambipolarity produced by internal plasma turbulence in a region must transfer the forces from the body to the  boundary of that region.  This is done when magnetic islands lock magnetic surfaces into a co-rotating state.  But this transfer can also be made by a viscosity-like force on the electrons.  

The form of the viscosity-like force on the electron fluid velocity is
\begin{eqnarray}
&& \vec{F}_v \equiv -  \vec{\nabla}\times(\nu\vec{\nabla}\times\vec{v}_e), \mbox{    which satisfies   } \hspace{0.3in} \label{F_v def} \\
&& \int \vec{F}_v d^3x = \oint (\nu\vec{\nabla}\times\vec{v}_e) \times d\vec{a}.
\end{eqnarray} 
The force exerted throughout a volume is transmitted to the bounding surface with $\nu(\vec{x},t)$ an effective viscosity coefficient.  Unlike the locking of magnetic islands,  the viscous force dissipates kinetic energy: 
\begin{eqnarray}
\int \vec{v}_e\cdot \vec{F}_v d^3x &=& -\int \nu (\vec{\nabla}\times\vec{v}_e)^2 d^3x \nonumber\\&& - \oint \nu (\vec{v}_e\times \vec{\nabla}\times\vec{v}_e)\cdot d\vec{a}.
\end{eqnarray}

The viscosity coefficient becomes non-zero when the islands become sufficiently broad that islands that are centered on different surfaces overlap each other.

Island overlap is the heuristic Chirikov criterion \color{black} \cite{Chirikov}\color{black}  for the transition to magnetic field line chaos.  An individual field line can cross an entire chaotic region.  Nevertheless, the cores of some islands remain, and the transition is more complicated than is often assumed \color{black} \cite{Meiss}\color{black}.  

A central concept is the most irrational rotational transform \color{black} $\iota \equiv 1/q$, where $q$ is the safety factor\color{black}.  This is the transform that requires the highest integer $M$ to represent $\iota$ within a given error $\epsilon$ by the ratio of two integers $N/M$.  A topological transition occurs when the magnetic surface with the most irrational transform between two island chains breaks by forming a cantorus.  A cantorus is like an irrational magnetic surface except it has pairs of holes, called turnstiles, through which magnetic field lines can pass---one hole of the pair has outgoing lines and the other an equal flux of incoming lines.  A review is given in \cite{Chaos:Boozer}, and the discussion in Section \ref{sec:islands}  explains why the turnstiles form at the points where the neighboring unperturbed magnetic surfaces in a stellarator are closest together.

The electrons can pass through the turnstiles and move move across the surfaces of constant  electron pressure $p_e$.  An electric potential must arise along these magnetic field lines to preserve quasi-neutrality, $\vec{B}\cdot\vec{\nabla}\Phi = (\vec{B}\cdot\vec{\nabla}p_e)/en_e$ to ensure the electron number density $n_e$ equals that of the ions.  

Once the nested magnetic surfaces are replaced by cantori, the field lines trajectories have extreme complexity in space, which implies that a slowly varying electric potential along the magnetic field lines produces rapid variations in $\Phi$ across the magnetic field lines.  The resulting $\vec{E}\times\vec{B}$ flows give transport comparable in magnitude to Bohm diffusion \cite{Boozer:2024}, $D_B=T/eB$, which equals $\rho_sC_s$.  The viscosity coefficient can reach the value
\begin{equation}
\nu \approx m_i n_e D_B.
\end{equation}
The ion mass arises from the requirement that the ions participate in the $\vec{E}\times\vec{B}$ flows needed to preserve quasi-neutrality.

\end{enumerate}
\color{black}

The viscosity-like force that damps gradients in the electron rotation velocity requires a sufficiently high plasma beta, $\beta>\beta_{ch}$, to produce magnetic island overlap over a large fraction of the plasma radius.  Although both the viscosity-like force and the island-locking force are present for $\beta>\beta_{ch}$, the island locking force will be found to be the larger force produced by what is nominally electrostatic microturbulence is either tokamaks or stellarators.
\color{black}

The major conclusions of the paper are discussed in Section \ref{sec:discussion}.



\section{Perturbed magnetic field due to microturbulence \label{sec:B-pert} }

The effect of microturbulence on the magnetic field is easier to understand when the velocity of the magnetic field lines $\vec{u}_\bot$ is distinguished from the mass-flow velocity $\vec{v}$ of the plasma.  In a torus with $\varphi$ the toroidal angle, the field-line velocity can be defined using the representation of an arbitrary vector $\vec{E}$ in terms of another arbitrary vector $\vec{B}$,  Equation (26) in \cite{Boozer:RMP},
\begin{equation}
\vec{E}+\vec{u}_\bot \times \vec{B} = - \vec{\nabla}\Phi + V_\ell \vec{\nabla}\frac{\varphi}{2\pi}, \label{E-exp}
\end{equation}
which is a mathematical identity.  This equation can be inserted into Faraday's Law to obtain an equation with general validity,
\begin{eqnarray}
\frac{\partial \vec{B}}{\partial t} = \vec{\nabla} \times (\vec{u}_\bot \times \vec{B}) + \frac{\vec{\nabla}V_\ell \times \vec{\nabla}\varphi }{2\pi}. \label{Ad-diff}
\end{eqnarray}
When the term $(\vec{\nabla}V_\ell \times \vec{\nabla}\varphi)/2\pi$ involves second derivatives of $\vec{B}$ with respect to position then it is diffusive.  This is the case when there is a term $\eta\vec{j}$ in the expression for the electric field, \color{black} which diffuses magnetic field lines with $\eta/\mu_0$ the diffusion coefficient. \color{black} Equation (\ref{Ad-diff}) is then an advection-diffusion equation.  In 1984, Aref showed  \cite{Aref:1984} solutions to the advection-diffusion equations have an unexpected tendency to give rapid mixing even as diffusion coefficient becomes arbitrarily small.  All that is required is that the advective velocity be chaotic.

\begin{figure}
\centerline{ \includegraphics[width=3.2 in]{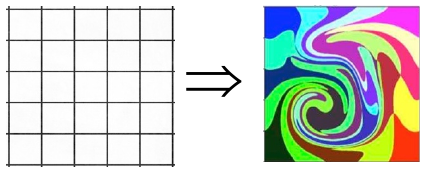}}
\caption{ A magnetic field $\vec{B}(\vec{x},t)$ can be thought of as consisting of tubes of magnetic flux by placing a gridded surface across the field.  Each tube is defined by the magnetic field lines that pass through the perimeters of the grid cells.  When the field is chaotic, the perimeter of each cell becomes exponentially longer when the grid is replotted after each line on the perimeters is followed for a distance $\ell$.   But, each cell contains exactly the same field lines and has precisely the same neighboring cells.  When the magnetic field is evolving ideally with a chaotic velocity $\vec{u}_\bot$, a similar distortion of the grid occurs when the grid is replotted using the location of each line on the perimeters after a time $t$.  The figure shows the distortion of a $5\times5$ array.   This is Figure 1 of Boozer, Phys. Plasmas \textbf{32}, 052106 (2025).  The distorted grid is part of Figure 5 of Y.-M. Huang and A. Bhattacharjee, Phys. Plasmas 29, 122902 (2022), which was based on a chaotic evolution defined by A. H. Boozer and T. Elder, Phys. Plasmas \textbf{28}, 062303 (2021).  Boozer and Elder illustrated distortions of ideally evolving flux tubes up to a factor $\sim 10^7$.  } 
\label{fig:tubes}
\end{figure}

 A magnetic field line flow $\vec{u}_\bot$ driven by turbulence is chaotic, which means neighboring streamlines of the flow separate exponentially in time with a timescale $\tau_u$. What many find surprising is that magnetic surfaces will break on a timescale only an order of magnitude longer than $\tau_u$ no matter how small the resistive diffusion $\eta/\mu_0$ may be.  Resistivity enters the reconnection timescale only logarithmically \cite{Boozer:surface, Chaos:Boozer}.  Figure \ref{fig:tubes} illustrates how chaos exponentially enhances the effect of $\eta/\mu_0$ diffusion on reconnection.  

Although chaotic flows can rapidly break magnetic surfaces, the timescale for the re-formation of magnetic surfaces is not shortened by a chaotic flow.  A resistive timescale is required---not the logarithm of the resistive timescale as  in the breaking.  

The lack of time reversibility in mixing is well known in everyday life.  Ingredients in soup are mixed by stirring, but they cannot then be separated by stirring in the opposite direction.  The time-reversibility of solutions to an advection-diffusion equation are exponentially sensitive to the diffusion, which rules out flutter in the topological features of the magnetic field lines when resistive diffusion is small. \color{black} Although the exponential sensitivity of time reversibility in chaotic flows may seem obvious, its fundamental importance to the theory of magnetic reconnection is largely ignored \cite{Browning}. \color{black}

A slab-model, in which a constant magnetic field $B_z\hat{z}$ is subjected to a perturbation $\tilde{B}_x\hat{x}$, illustrates the relationship between the magnetic field line velocity $\tilde{u}_x\hat{x}$ and the plasma velocity $\tilde{v}_x\hat{x} + \tilde{v}_z\hat{z}$.  Equation (\ref{Ad-diff}), with an ideal perturbation, $\vec{\nabla}V_\ell=0$,  implies $ \partial_t \tilde{B}_x = B_z\partial_z \tilde{u}_x$, so $ \partial_z \partial_t \tilde{B}_x =  B_z\partial_z^2 \tilde{u}_x$.  Ampere's Law implies $ \partial_z \tilde{B}_x = \mu_0 j_y$ and force balance gives $j_y B_z = m_i n \partial_t \tilde{v}_x$, which leads to the equation $ \partial_z \partial_t \tilde{B}_x =  \frac{\mu_0 m_i n}{B_z}  \partial_t^2 \tilde{v}_x.$   Equating the two expressions for $ \partial_z \partial_t \tilde{B}_x$,
 \begin{eqnarray}
  \frac{\partial^2 \tilde{u}_x}{\partial z^2} &=&  \frac{\mu_0 m_i n}{B_z^2} \frac{\partial^2 \tilde{v}_x}{\partial t^2}\\
 &=&  \frac{1}{V_A^2} \frac{\partial^2 \tilde{v}_x}{\partial t^2}.
 \end{eqnarray}
  When the phase velocity of $\tilde{v}_x$ along $z$ equals the Alfv\'en speed, $\tilde{u}_x=\tilde{v}_x$, and the whole motion of the plasma is due to the motion of the magnetic field lines, which means an Alfv\'en wave.  When the phase velocity along $z$ is approximately the \color{black} the speed of sound $C_s=\sqrt{T/m_i}$, then $\tilde{u}_x = (C_s^2/V_A^2)\tilde{v}_x$.  The ratio $C_s^2/V_A^2 \sim \beta$, \color{black} and the magnetic field lines move little compared to the plasma motion.  Nevertheless, the velocity of the magnetic field lines is turbulent and, therefore, chaotic, which implies magnetic reconnection that breaks the magnetic surfaces will quickly occur \cite{Boozer:surface}, regardless of how small non-ideal effects represented by the loop voltage may be.

When potential fluctuations $\tilde{\Phi}$ have a phase velocity along the field lines $\approx C_s$, there is cross magnetic surface magnetic fluctuation 
\begin{eqnarray} 
\frac{\tilde{B}}{B} &\approx& \left(\frac{C_s}{V_A}\right)^2 \frac{e\tilde{\Phi}}{T} \\
& \approx& \beta \frac{ \tilde{e\Phi}}{T }  \label{Phi/T}\\
& \approx& \beta \frac{ \tilde{\rho}_s}{a } \label{Phi-fluc}
\end{eqnarray}
since the typical microturbulent fluctuation amplitude is $e\tilde{\Phi}/T\approx\rho_s/a$ with $a$ the plasma radius.


\section{Properties of the magnetic islands \label{sec:islands} }

Due to the singularity in the current density that arises on resonant rational surfaces even when the magnetic field evolves ideally \cite{Current-sing}, islands of width $\Delta_{mn}$ quickly arise on each magnetic surface that resonates with the magnetic fluctuations produced by the microturbulence.  The singularity implies quick island formation, but no similar effect allows an island to disappear.  Microtubulence does not cause islands to flutter into and out of existence.

To simplify the arguments, circular magnetic surfaces will be assumed with the radius of those surfaces assumed to be related to the enclosed toroidal magnetic flux by $\psi_t=B_\varphi \pi r^2$.  Actually, the magnetic surfaces in stellarators are strongly shaped with positions on the surfaces given by $\vec{x}(r,\theta,\varphi)$, with $\theta$ the poloidal, and $\varphi$ the toroidal angle.  Neither $|\partial \vec{x}/\partial r|=1$ nor $|\vec{\nabla}r |=1$, as would be the case withe circular surfaces.  The resulting geometric factors are given in the equations of central importance. 

A Fourier term in the magnetic perturbation produced by the turbulence is $\tilde{b}_{nm}\sin(n\varphi-m\theta) \equiv \tilde{B}_r/B$, where the fluctuation $\tilde{B}_r$ is perpendicular to the magnetic surfaces.  The rotational transform $\iota(r) =RB_\theta(r)/rB$ in the unperturbed magnetic field.  The magnetic perturbation is resonant at $\iota(r_{mn})=n/m$, and the half-width of the resulting island is
\begin{eqnarray}
\Delta_{mn} = \sqrt{\frac{4 r_{mn} \tilde{b}_{nm}}{m \frac{d\iota}{dr}} }.
\end{eqnarray}
A derivation including all geometric effects is given in \cite{Boozer:RMP} and the result is given in Equation (15) of that reference.  Including geometric effects, the normal magnetic field $B_r$ should include a factor $|\vec{\nabla}r|$.

The distance $D_{mn}$ between rational surfaces is also important.  $D_{mn}$ is defined by $\iota(r_{mn} + D_{mn})=\iota(r_{m-1,n})$.  When $m>>1$,
\begin{eqnarray} 
D_{mn} &=& \frac{1}{s_I} \frac{r_{mn}}{m}\Big|\frac{\partial \vec{x}}{\partial r}\Big| \\
s_{mn} &\equiv &\Big(\frac{d\ln(\iota)}{d\ln(r)}\Big)_{r_{mn}}
\end{eqnarray}
is the shear at the resonant surface

When $2\Delta_{mn} >D_{mn}$ within a region of the plasma, islands on different rational surfaces overlap.  According to the heuristic Chirikov criterion \color{black}\cite{Chirikov}, \color{black} overlap implies magnetic field lines exist that have a separation from neighboring lines that have an exponential dependence on the distance along the lines and  come arbitrary close to every point in a volume within that region.  This is the chaotic magnetic field line limit.   When $\Delta_{mn} << D_{mn}$ perfect magnetic surfaces exist in the annular regions between islands.  The existence of magnetic surfaces bounds the excursions of magnetic field lines.  The transition between these two limits is subtle involving cantori and turnstiles \color{black} \cite{Meiss} \color{black} and is reviewed in Reference \cite{Chaos:Boozer}.  

Using Equation (\ref{Phi-fluc}) to define $\tilde{b}_{nm}$ and letting $\tilde{b}_{nm}^{ch}$ be the perturbation required to satisfy the Chirikov overlap criterion, then island overlap or not depends on whether
\begin{eqnarray}
f_{ch} &\equiv& \frac{\tilde{b}_{nm}}{\tilde{b}_{nm}^{ch}}  \\
&\approx& 16\beta \iota_{mn} s_{mn} \frac{|\vec{\nabla}r|}{\Big|\frac{\partial \vec{x}}{\partial r}\Big|^2} \label{Chirikov}
\end{eqnarray}
 is larger or smaller than one.  The geometric factor at the end of Equation (\ref{Chirikov}) \color{black} depends inversely on the third power of the local separation between  neighboring magnetic surfaces\color{black}.     These are the locations of the turnstiles that break the remaining spatial confinement of the magnetic field lines.

Equation (\ref{Chirikov}) implies the plasma $\beta$ must exceed a critical value $\beta_{ch}$ for the magnetic field lines to have large scale chaos,
\begin{equation}
\beta_{ch} \approx \frac{1}{16 r_{mn} (d\iota/dr)_{mn}} \left( \frac{ \left| \frac{\partial\vec{x}}{\partial r} \right|_{mn}^2}{\left|\vec{\nabla}r\right|_{mn}} \right)_{min}, \label{beta_ch}
\end{equation}
where the subscript ``min" means the expression should be evaluated at the point at which neighboring magnetic surfaces have their closest approach.


\section{Force that gives non-ambipolar transport \label{sec:NA} }

The momentum conserving force balance equations for electrons and ions are
\begin{eqnarray}
m_en_e \frac{d\vec{v_e}}{dt} &=&- en_e (\vec{E} + \vec{v}_e \times \vec{B}) -\vec{\nabla}p_e - \vec{F}_a + \vec{F}_v; \nonumber\\ && \\
m_in_e \frac{d\vec{v}}{dt} &=& en_e (\vec{E} + \vec{v} \times \vec{B}) -\vec{\nabla}p_i +\vec{F}_a+ \vec{F}_{na}, \nonumber\\ &&
 \end{eqnarray}
 where the mass-flow speed is identified with the ion velocity.   The electrons and ions have the same number density, $n_e$.  The ions have an intrinsic non-ambipolar drag force $\vec{F}_{na}$, which will be derived below in the approximation of circular magnetic surfaces.  \color{black} The ion-electron interaction force $\vec{F}_a$ enters the electron and ion equations with opposite signs, which gives ambipolar transport.  $\vec{F}_v$ is the viscosity-like force acting on the electrons. \color{black}
 
 Since  $\vec{j}=en(\vec{v} - \vec{v}_e)$, when the the electrons are assumed to be massless,  the sum of the two force-balance equations is
\begin{eqnarray}
m_i n \frac{d\vec{v}}{dt} = \vec{j}\times\vec{B} - \vec{\nabla}p + \vec{F}_{na} + \vec{F}_v. \label{fluid-force}
\end{eqnarray}


When the magnetic surfaces are nearly circular, $(r,\theta,\varphi)$ orthogonal coordinates can be used, which are $(r,\theta,z)$ cylindrical coordinates with $\varphi \equiv z/R$.  The $z$ coordinate is periodic with a period $2\pi R$.  The minor radius $r$ is assumed to be very small compared to the major radius $R$.  The flows associated with equilibrium balance are in the $\hat{\theta}\equiv r \vec{\nabla}\theta$ direction.

Equilibrium force balance occurs on a fast timescale compared to rotation changes and requires $\vec{j}\times\vec{B} = \vec{\nabla}p(r)$.  The implication is that $en_e(v-v_e)\hat{\theta}\times B_\varphi \hat{\varphi} = (dp/dr) \hat{r}$.  The ion and electron flows are in the $\hat{\theta}$ direction and obey
\begin{eqnarray}
v&=& v_e +  v_p\\
v_p &\equiv&  \frac{dp/dr}{en_e B_\varphi}\\
|v_p|&\approx& \frac{T}{aeB_\varphi}\approx \frac{\rho_s }{a} C_s
\end{eqnarray} 
where $v_p$ is assumed to be a constant in time.

 The ions intrinsic non-ambipolar drag force is
 \begin{eqnarray}
  \vec{F}_{na} &=& j_r \hat{r} \times B_\varphi\hat{\varphi} \\
  &=&- j_rB_{\varphi}\hat{\theta}.\\
 | j_r| &=& en_e \frac{a}{\tau_p^{na}} \left| \frac{v}{v_p}\right|
  \end{eqnarray}
  The non-ambipolar particle confinement time $\tau_p^{na}= a^2 / f_{na} D_{gb}$, where the gyro-Bohm diffusion coefficient $D_{gb}\equiv (\rho_s/a)(T/eB)=\rho_s^2 C_s/a$ gives the typical empirical level of energy transport, $f_{na}$ is the dimensionless relative strength of the non-ambipolar transport compared to the gyro-Bohm transport, and $B=B_{\varphi}$.  Consequently, $\tau_p^{na}=a^3/(\rho_s^2 C_s f_{na})$.
  
The maximum non-ambipolar force has $|v/v_p|=1$, which means the ions are carrying the diamagnetic current that balances the gradient in the pressure, $p$, and the non-ambipolar force has the characteristic value
  \begin{eqnarray}
  F_{na} &=& f_{na} \frac{\rho_s^2}{a^2} \ en_e C_s B\\
   &=&f_{na} \frac{\rho_s}{a} \frac{p}{a}.  \label{F_na}
  \end{eqnarray}
 
 
 \section{Magnitude of electron force with separated islands \label{sec:separated}}

When neighboring islands are separated, $f_{ch}<1$, they exert a non-dissipative force on each other that is  sinusoidal in the relative phases given by their two $m\theta-n\varphi$ dependencies.  This force is given by $\vec{j}_i \times\vec{B}_{i'}$, where  $\vec{j}_i$ is the current density that produces one island and $\vec{B}_{i'}$ is the magnetic perturbation associated the other island at the location of $\vec{j}_i$.

In the limit of thin islands, the current density $\vec{j}_i$ can be approximated as a delta function on the rational surface. The magnetic perturbation $\vec{j}_i$ produces is then otherwise curl and divergence free:
\begin{eqnarray}
\frac{B_r}{B}&=& \tilde{b}_{mn}  \exp\Big( - m \frac{|r-r_{mn}|}{r_{mn}}\Big) \times \nonumber\\&& \hspace{1.3in}\sin(m\theta-n\varphi)  \\
\frac{\tilde{B}_\theta}{B} &=& \mp\tilde{b}_{mn}  \exp\Big( - m \frac{|r-r_{mn}|}{r_{mn}}\Big)\times \nonumber\\&& \hspace{1.3in} \cos(m\theta-n\varphi),
\end{eqnarray}
where the $\mp \equiv - (r-r_{mn})/|r-r_{mn}|)$.

The current flowing on the rational surface is
\begin{eqnarray}
&& j_z = J_I\delta(r-r_{mn}) \cos(m\theta-n\varphi)  \mbox{   with  } \\
&&J_I  \cos(m\theta-n\varphi) = \frac{m[B_\theta]}{r_{mn}},  \mbox{   so  }   \\
&&J_I = - 2 \tilde{b}_{mn}B \frac{m}{r_{mn}}.
\end{eqnarray}
$[B_\theta]$ is the jump in the value of $B_\theta$ across the resonant surface.  The distance $r_{mn}/m\approx\rho_s$.

The force per unit volume exerted on the rational surface $(m,n)$ by the surface $(m',n')$ is approximately $(j_z)_{mn} (\delta B_r)_{m'n'} \exp(-D_{mn}/\rho_s)$.  Let 
\begin{eqnarray}
\delta_{mn}&\equiv& \frac{D_{mn}}{\rho_s} \\
 &\approx& \frac{\Big|\frac{\partial \vec{x}}{\partial r}\Big|_{r_{mn}} }{s_{mn}}.
\end{eqnarray}
The normalized separation $\delta_{mn}$ is small where the two unperturbed magnetic surfaces are close together---as one would expect.

  Integrating across the delta function, the force per unit area in the $\theta$ direction, which tends to lock the two surfaces together, is
\begin{eqnarray}
&& F_\theta = - F_{\ell}\frac{\cos(m\theta-n\varphi) \sin(m'\theta-n'\varphi)}{2}.
\end{eqnarray}
Ignoring numerical factors,
\begin{eqnarray}
F_\ell &\approx& \tilde{b}_{mn}\tilde{b}_{m'n'}\frac{B^2}{\mu_0} \frac{m}{r_{mn}}  \exp(-\delta_{mn}) \\
&\approx& \beta^2 \frac{\rho_s}{r_{mn}^2} \frac{B^2}{\mu_0} \exp(-\delta_{mn}) \\
& \approx& \beta \frac{\rho_s}{r_{mn}} \frac{p}{r_{mn}} \exp(-\delta_{mn}).
\end{eqnarray} 
$F_{\ell}$ is a force that tends to lock neighboring islands in a certain phase relation.

Comparing the locking force $F_{\ell}$ to the intrinsic non-ambipolar force $F_{na}$ of Equation (\ref{F_na}) 
\begin{eqnarray}
&& \frac{F_\ell}{F_{na}} \approx \frac{\beta}{f_{na}} \exp(-\delta_{mn}).
\end{eqnarray}  

Plasma beta values tend to be far larger than $f_{na}$ in cases of primary interest, so the islands will lock unless $\delta_{mn}\gtrsim3$.

\section{Effect on electrons with overlapping islands}

When the Chirikov factor $f_{ch}$, Equation (\ref{Chirikov}) exceeds unity magnetic field lines and electrons can make large radial excursions by passing through turnstiles.  

As noted in the Introduction, to enforce quasi-neutrality a variation in the electric potential $\Phi$ along the field lines is required.   The quasi-neutrality potential is slowly varying with distance $\ell$ along field lines.  Nevertheless, the spatial complexity of the field line trajectories and the exponential dependence of the separations of neighboring lines on $\ell$ makes $\vec{\nabla}_\bot \Phi$, the gradient perpendicular to the lines, extremely large.  The resulting $\vec{E}\times \vec{B}$ plasma flows and radial transport give a transport that tends to be comparable to Bohm diffusion, $D_B\equiv T/eB\approx \rho_sC_s$ \cite{Boozer:2024}.

 The radial motion of the electrons changes the radial electric field and thereby the $\vec{E}\times\vec{B}$ flow of the ions.  The implication is that 
\begin{equation}
\nu \approx m_i n _e D_B. 
\end{equation}   

Since the characteristic flow speed on the electrons carrying the confining current is $\rho_sC_s/a$, the characteristic magnitude of the viscous force is 
\begin{eqnarray}
F_v &\sim& \frac{\nu (\rho_sC_s/a)}{a^2}\\
& \approx&  m_i n_e \frac{(\rho_sC_s)^2}{a^3} \approx n_e T \frac{\rho_s^2}{a^3} \\
& \approx& \frac{p}{a} \frac{\rho_s^2}{a^2}.
\end{eqnarray}

The ratio of the viscous force to the intrinsic non-ambipolar force $F_{na}$ of Equation (\ref{F_na}) is
\begin{eqnarray}
\frac{F_v}{F_{na}} &\approx& \frac{ \frac{\rho_s}{a}}{ f_{na} }.
\end{eqnarray}
For the viscous force to change the electric potential, the non-ambipolar fraction must be sufficiently small that $f_{na}< \rho_s/a$.

The force associated with the phase locking of islands does not go away when the Chirikov factor is larger than unity; the cores of islands persist.  But, the current density is the islands presumably spreads over the width of the island, so the $\exp(-\delta_{mn})$ factor is presumably near unity.  The ratio
\begin{eqnarray}
\frac{F_v}{F_\ell} &\approx& \frac{\frac{\rho_s}{a} }{ \beta }, 
\end{eqnarray}
which implies the viscous force makes a subdominant contribution to stopping the electrons.

  \section{Discussion \label{sec:discussion} }
  
  Arguments for the ambipolarity of micturbulence are based on the assumption that magnetic surfaces remain intact, but this assumption untenable.  At any non-zero plasma beta, the magnetic fields associated with what are considered electrostatic instabilities produce islands that split the magnetic surfaces with which they resonate. 
  
 Even an ideally evolving magnetic perturbation drives a delta-function current density \cite{Current-sing} on any magnetic surface on which the rotational transform $\iota$ equals $n/m$ with $m$ a poloidal $n$ a toroidal mode number of the perturbation.  The singularity of the delta-function current density causes the immediate opening of an island.  There is no singularity that allows a rapid closure of an island.  The island can rotate as the electron fluid rotates, but the islands cannot flutter into and out of existence.
  
  Islands on different rational surfaces exert a non-dissipative force on each other, which has a sinusoidal dependence on the relative phases of the neighboring islands.  This force can be sufficiently strong to lock the islands into a common rotation rate even at a low value of the plasma $\beta$.  This locking impedes the electrons carrying the current that supports the pressure gradient, which tends to produce a large scale radial electric field that expels impurities from the plasma.
  
 The width of the islands depends on the plasma $\beta$.  When $\beta$ exceeds a value $\beta_{ch}$, which is given in Equation (\ref{beta_ch}), islands from different rational surfaces overlap.  Some magnetic field lines then become radially unbounded.  The existence of radially unbounded magnetic field lines has a large effect on the electrons.  In particular, the electrons develop a viscosity-like force that damps gradients in their rotation velocity.  Remarkably, the viscous damping of rotation gradients  produces a smaller force on the electrons than does the island locking force.  Both forces are present when $\beta>\beta_{ch}.$
 
 The arguments given in this paper are too heuristic for the reliability of the numerical values.  The hope is that they will inspire numerical simulations that will allow quantitive predictions. \\

\section*{Acknowledgements}

The author thanks the reviewer of an earlier version for asking a question on whether the breaking of surfaces is on the equilibrium or on the mictrotubulent scale.  That version considered only the viscosity-like damping of the electron rotation.  While considering an answer, the author recognized that the island locking forces, which arise at $\beta$ values too low for the viscosity-like damping, actually produce a larger force to block the free rotation of the electrons under simple assumptions at all values of $\beta.$

This work was supported in part by the U.S. Department of Energy, Office of Science under Award No. DE-AC02-09CH11466).



 \end{document}